\documentclass{article}
\usepackage{spconf,amsmath,graphicx}
\usepackage{booktabs}
\usepackage{caption}
\usepackage{pifont}
\newcommand{\cmark}{\ding{51}}%
\newcommand{\xmark}{\ding{55}}%

\title{Improving Noise Robustness of Contrastive Speech Representation Learning with Speech Reconstruction}
\name{\begin{tabular}{c}Heming Wang$^{1*}$\thanks{*Work done during an internship at Microsoft.}, Yao Qian$^2$, Xiaofei Wang$^2$, Yiming Wang$^2$, Chengyi Wang$^2$, \\
Shujie Liu$^2$, Takuya Yoshioka$^2$, Jinyu Li$^2$, DeLiang Wang$^{1}$ \end{tabular} }

\address{
$^1$The Ohio State University, USA, $^2$Microsoft Corporation\\
\texttt{\footnotesize wang.11401@osu.edu, \{yaoqian,xiaofewa,yimingwang,t-chewang,shujliu,tayoshio,jinyli\}@microsoft.com,}\\
{\texttt{\footnotesize dwang@cse.ohio-state.edu}}
}

\begin{document}
\ninept
\maketitle
\begin{abstract}
Noise robustness is essential for deploying automatic speech recognition (ASR) systems in real-world environments. One way to reduce the effect of noise interference is to employ a preprocessing module that conducts speech enhancement, and then feed the enhanced speech to an ASR backend. In this work, instead of suppressing background noise with a conventional cascaded pipeline, we employ a noise-robust representation learned by a refined self-supervised framework for noisy speech recognition. We propose to combine a reconstruction module with contrastive learning and perform multi-task continual pre-training on noisy data. The reconstruction module is used for auxiliary learning to improve the noise robustness of the learned representation and thus is not required during inference. Experiments demonstrate the effectiveness of our proposed method. Our model substantially reduces the word error rate (WER) for the synthesized noisy LibriSpeech test sets, and yields around 4.1/7.5\% WER reduction on noisy clean/other test sets compared to data augmentation. For the real-world noisy speech from the CHiME-4 challenge (1-channel track), we have obtained the state of the art ASR performance without any denoising front-end. Moreover, we achieve comparable performance to the best supervised approach reported with only 16\% of labeled data.


\end{abstract}
\section{Introduction}
\label{sec:intro}
Automatic speech recognition (ASR) has seen rapid progress recently by employing deep learning techniques. However, speech in real-world environments usually contains background noise, which significantly degrades the ASR performance \cite{haeb2019speech}.
Early works attempted to tackle this problem with a network that performs pre-processing, as deep neural network (DNN)-based speech enhancement approaches have demonstrated superior performance in suppressing background noie \cite{wang2018overview,hu2020dccrn}.
During inference, these networks are adopted as front-end modules and the enhanced speech is fed to acoustic model backends.

Although these networks effectively removed the background noise and improved speech intelligibility \cite{wang2018overview,hu2020dccrn,wang2020complex}, the ASR performance of the noise-suppressed speech may be degraded, caused by the nonlinear distortion brought by DNN. 
A common strategy is to jointly train the enhancement network and DNN-based acoustic models, such that the distortions brought by the pre-processors are alleviated \cite{meng2017deep, eskimez2021human, subramanian2019speech, watanabe2020ESPupdate,wang2021exploring}. 
However, this may introduce more complex network architectures and training strategies.

Self-supervised learning leverages high-resource unlabeled data, and learns a speech representation beneficial for downstream tasks like speech recognition. It is commonly believed that through large-scale self-supervised learning,
a speech representation more effective than conventional features like waveforms and Mel-spectrograms can be obtained.
Even with a limited amount of labeled data, experiments show that these frameworks achieve strong ASR performance.
Self-supervised architectures can be categorized into two kinds. One is generative/predictive learning \cite{alvarez2019pase, chung2019apc, liu2020mockingjay, ling2020bertphone}, which aims to generate/predict the original input using information extracted from past time steps or masked inputs.
The other uses contrastive learning \cite{baevski2020wav2vec, oord2018cpc, baevski2019vq,kawakami2020learning,wang2021unispeech} to obtain high-level representations by solving a contrastive task in a latent embedding space.

In this paper, instead of following the conventional supervised paradigm, we propose a novel self-supervised approach to address the robust ASR problem. Specifically, we combine a reconstruction module with the contrastive learning framework of wav2vec 2.0 \cite{baevski2020wav2vec} to improve the noise robustness of learned speech representations in the pre-training stage.
There were also recent self-supervised studies that address the robust ASR problem. Ravanelli et al. \cite{ravanelli2020paseplus} performed supervised generative/predictive tasks with 12 workers (small DNNs), and conducted online contamination on clean input speech. Sadhu et al. \cite{sadhu2021wav2vecc} combined wav2vec 2.0 and VQ-VAE, and attempted to reproduce the quantized representations from latent speech embeddings. Different from \cite{ravanelli2020paseplus}, we optimize both reconstructive and contrastive losses through continual pre-training with limited noisy speech data. Additionally, Sadhu et al. \cite{sadhu2021wav2vecc} did not explicitly incorporate the denoising process in the self-supervised training infrastructure, but mainly focused on improving the codebook utilization ratio of wav2vec 2.0. The main contribution of this paper is three-fold:
\begin{enumerate}
    \item We incorporate reconstruction learning into the contrastive learning framework of wav2vec 2.0, and conduct continual pre-training to explicitly reconstruct the clean speech from the noisy input.
    \item Experiments show that our proposed approach is capable of improving the noise robustness of learned representations. Our approach yields significant WER improvement for synthetic noisy speech, and at the same time maintains the performance for clean speech. 
    \item Without involving front-end preprocessing, our approach can match the state of the art performance of supervised approaches with only 16\% of labeled training data on real-world noisy speech of CHiME-4.
\end{enumerate}




\begin{figure*}[ht]
\includegraphics[width=0.8\textwidth]{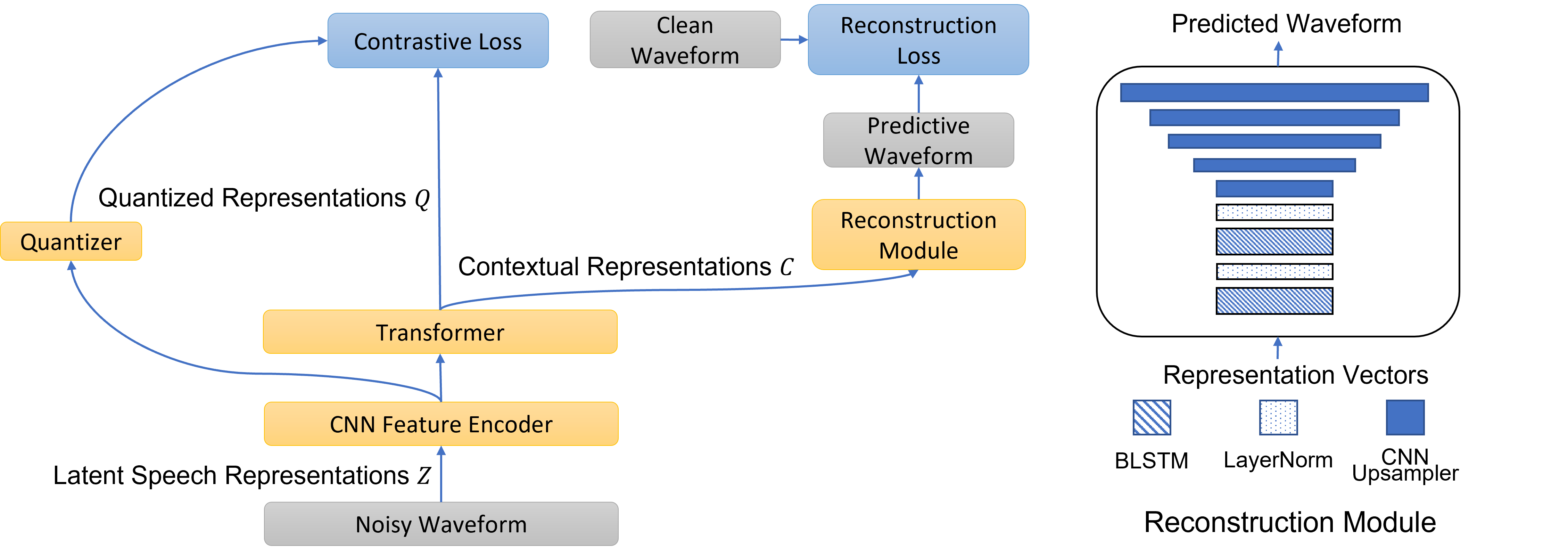}
\centering
\caption{The left diagram illustrates the proposed self-supervised framework. On top of the contextual representation, we apply a reconstruction module to enforce it to carry high-level information for the corresponding clean speech. The right diagram displays the structure of the reconstruction module, consisting of a two-layer BLSTM and a CNN upsampler to guarantee the prediction has the same length as the input.}
\label{fig:fairseqrc}
\end{figure*}


\section{Network Architecture}
\label{sec:network}

\subsection{Wav2vec 2.0}
Our method is based on the contrastive learning framework of wav2vec 2.0 \cite{baevski2020wav2vec}.
In wav2vec 2.0, a convolutional neural network (CNN) feature encoder is applied to the input speech to extract latent speech representations $Z$.
Then representations $Z$ are discretized using a Gumbel Vector quantizer to generate quantized embedding $Q$.
Finally, we employ masks on the latent representation vectors, which are then passed to a transformer-based context network to generate contextual representations $C$ that will be used for downstream tasks.
A contrastive loss is calculated on contextual representations and quantized embeddings, such that the context network can identify the true quantized representations (sampled from the current masked time step) from a set of distractors (quantized representations sampled from other time steps within the same utterance).
\subsection{Proposed Architecture}
As illustrated in Fig. \ref{fig:fairseqrc}, a reconstruction module is added atop the contextual representations $C$ in the network.
The network takes in a noisy waveform as the input and passes the contextual representations to the reconstruction module to predict the corresponding clean waveform. 
During training, we conduct multi-task learning which optimizes the network to reconstruct the clean speech, and at the same time, solves the contrastive task by distinguishing the positive samples from the negative samples.
This way we enforce the representation vectors $C$ to explicitly carry sufficient information to represent the corresponding clean speech. 
The design of the reconstruction module is inspired by the convolutional recurrent network (CRN) that has been successfully applied in speech enhancement studies \cite{tan2018complex,hu2020dccrn}. 
CRN is composed of a CNN encoder, a recurrent neural network-based bottleneck module for modeling the temporal dependencies, and a CNN decoder. We regard the modules before contextual representations as the encoder and construct the reconstruction module by using a two-layer bidirectional long short-term memory (BLSTM) network followed by a CNN decoder. In the bottleneck, we perform layer normalization after each BLSTM layer. The CNN decoder is designed to be symmetric to the CNN feature extractor of wav2vec 2.0 such that we fully recover the original shape of the input waveform. 
Note that reconstruction is only performed in the pre-training stage, and it is not required in fine-tuning and inference.



\subsection{Loss}
As with the original wav2vec 2.0 loss design, a contrastive loss $L_c$ is calculated based on context representations $C$ and quantized embeddings $Q$. Given a context vector $c_t$ at the current masked time step $t$, the model is trained to identify the true sample $q_t$, within a set of $K$ negative samples {$\bar{q}_1,\bar{q}_2, ...,\bar{q}_K$} that are uniformly sampled from other time steps of the same utterance.
\begin{equation}
    L_c = -\log \frac{\exp(c_t, q_t) / \tau} {\exp(c_t, q_t) + \sum_{i=1}^{K} \exp(sim(c_t, \bar{q}_i)/\tau)}, 
\end{equation}
where $\tau$ is the temperature, and the cosine similarity is employed as the similarity function $sim(\cdot)$.
In addition, a diversity loss $L_d$ is employed to encourage the equal usage of all entries within the codebooks of the quantizer.
\begin{equation}
    L_d = \frac{GV - \sum_{g=1}^{G} \exp(- \sum_{v=1}^V p_{g,v}\log p_{g,v})}{GV}, 
\end{equation}
where $G$ is the number of codebooks and $V$ is the number of entries in the cookbook.
Finally, we add the reconstruction loss $L_r$, which is obtained by calculating the mean absolute error between the predicted waveform $\hat{y}$, and the clean waveform $y$.
With the total number of time steps denoted as $T$, the reconstruction loss is defined as,
\begin{equation}
    L_r = \frac{1}{T} \sum_{t=1}^{T} |\hat{y}(t) - y(t)|. 
\end{equation}
Our training loss is composed of the three terms described above, i.e., 
\begin{equation}
    L = L_c + \lambda_1 L_d + \lambda_2 L_r.
\end{equation}
In our experiments, we empirically set $\lambda_1 = \lambda_2 = 0.1$ to stabilize training.

\section{Experiment}
\label{sec:experiment}

\subsection{Dataset}
\label{ssec:dataset}
To evaluate the performance of our proposed method, we have conducted experiments on both synthesized and real-world noisy speech corpora.
For the synthesized noisy speech, we generate training mixtures using the \texttt{train-clean-100} subset from LibriSpeech  \cite{panayotov2015librispeech}. The training noise is extracted from the DNS-challenge\footnote{https://github.com/microsoft/DNS-Challenge} by randomly selecting 20000 noise files, which has a duration of around 55 hours. 
For each utterance, we randomly mix it with a segment cut from the training noise at a signal-to-noise ratio (SNR) uniformly sampled from the \{5, 6, 7, ..., 20\} dB.
The results are reported for the \texttt{test-clean} and \texttt{test-other} subsets under both clean and noisy conditions. We evaluate the noisy speech by mixing testing utterances with unseen noise files extracted from the MUSAN corpus \cite{snyder2015musan} in the same SNR setting.

For real-world noisy speech, we perform experiments on the CHiME-4 challenge data \cite{vincent2016chime4}.
A six-channel distant microphone array and a close-talk microphone were used for recording by having people read text prompts from the Wall Street Journal (WSJ0) corpus \cite{paul1992wsj}.
The CHiME-4 dataset contains two types of noisy speech, real and simulated noisy speech. The real data was recorded under four challenging noisy environments, bus, cafe, pedestrian area, and street junction. The simulated data are generated by mixing clean utterances with background noise recorded in the four environments. The real portion of the one-channel track is selected for our evaluation.

\subsection{Continual Training}
On the simulated noisy LibriSpeech data, we employ the model that was pre-trained on the 960 hours of Librispeech data and with the BASE configuration \footnote{https://dl.fbaipublicfiles.com/fairseq/wav2vec/wav2vec\_small.pt}, which contains 12 transformer blocks, a model dimension of 768 for the contextual module and 8 attention heads.
We perform continuous pre-training with this model for 400k steps by using the 100-hour noisy Librispeech data described in Section \ref{ssec:dataset}.

For the realistic noisy data from the CHiME-4 challenge, we employ a different model that was pre-trained on the LibriVox corpus with the LARGE configuration\footnote{https://dl.fbaipublicfiles.com/fairseq/wav2vec/wav2vec\_vox\_new.pt}, which consists of 24 transformer blocks with a contextual module dimension of 1024 and 16 attention heads. We select this model because LibriVox \cite{panayotov2015librispeech} consists of 53.2k hours data, and in our experiments training on LibriVox exhibited better cross-corpus generalization performance. 
Continual training is performed on the simulated portion of the CHiME-4 training dataset for 400k steps.
We selected five out of the six channels of the simulated data (the second channel is not used because it does not face the speaker), and each microphone channel is regarded as an independent file.


\subsection{Fine-tuning}
We fine-tune the learned representations on labeled data, and generate character predictions by adding a linear adaptation layer on top of the transformer blocks.
Our models are trained using an Adam optimizer with a learning rate of 2e-4, and optimized by a CTC loss for 20k steps.

For the LibriSpeech corpus, the Libri-Light \cite{kahn2020librilight} 10-hour subset is selected for fine-tuning. Note that we do not add noise in the fine-tuning stage.
For real-world experiments, we perform fine-tuning on various percentages of labeled training data of the CHiME-4 challenge for 30k steps.
To generate the 50\% subset, for real noisy data, we first select the microphone channel that is most correlated with the other five. Using that as the reference, we select two other channels that have the top 2 correlation coefficients. For simulated noisy recordings, we select channels 1, 3, and 4. Similarly, a 16\% subset is generated by selecting the microphone channel that is most correlated with the other five in the real portion or channel 1 in the simulated portion. Finally a 5\% subset is created by randomly sampling utterances from the 16\% subset.

\vspace{-0.5em}

\subsection{Language Model and Decoding}
For LibriSpeech, we evaluate on the \texttt{test-clean} and \texttt{test-other} subsets. A 4-gram language model \cite{heafield2011kenlm} trained on LibriSpeech is applied with a beam size of 1500.
For CHiME-4 experiments, the results are evaluated using the \texttt{dev-real} and \texttt{eval-real} subsets of \texttt{isolated\_1ch\_track}. During decoding, we use a LSTM-based language model trained on the WSJ corpus \cite{wang2019espresso} that contains 65000 words, and apply it with a beam size of 500. 
Note that for both settings, we re-score the weights and the insertion penalty of language models on a development set with a Bayesian optimization\footnote{https://github.com/fmfn/BayesianOptimization}.


\section{Results and Analysis}
\label{sec:results}
\subsection{Results on Synthesized Noisy Speech}

\begin{table}[!tp]
\caption{Experimental results on synthesized noisy speech and clean speech  generated from LibriSpeech. We display the WER scores of fine-tuning on the 10hr Libri-Light data with and without applying a 4-gram language model.
}
\resizebox{0.5\textwidth}{!}{\begin{tabular}{@{}cccccc@{}}
\toprule
\textbf{Model}  & \textbf{LM}    & \multicolumn{2}{c}{\textbf{Clean}} & \multicolumn{2}{c}{\textbf{Noisy}} \\ \midrule
                &                & test-clean  & test-other  & test-clean  & test-other  \\ \midrule
Baseline        & \xmark         &   11.1      & 17.6        & 29.8        & 49.6        \\
                & \cmark         &   4.3       & 9.5         & 20.9        & 40.0        \\ \midrule
W/O RCModule    & \xmark         &   14.1      & 27.3        & 18.7        & 37.3        \\
                & \cmark         &   6.5       & 16.9        & 9.2         & 25.2        \\ \midrule
W/ RCModule     & \xmark         &   10.5      & 19.5        & 14.6        & 29.8        \\ 
                & \cmark         &   4.8      & 11.7       & \textbf{7.5}      & \textbf{20.2}        \\ \bottomrule
\end{tabular}}

\label{tbl:libri}
\vspace{-1.0em}

\end{table}

Table \ref{tbl:libri} presents the results for LibriSpeech subsets \texttt{test-clean} and \texttt{test-other} fine-tuning on the 10-hour labeled Libri-Light data.
As shown in the table, the baseline model where no continual training was performed, showed good performance for clean speech, but had a severe performance degradation when recognizing noisy speech.
Adding the continual training to the original model without a reconstruction module resulted in obvious WER reduction under noisy conditions; however, the ASR performance under clean conditions was hurt.
Specifically, a 3.2\% WER gain for \texttt{test-clean}, and 10.1\% for \texttt{test-other} were obtained (when no language model was applied). 
This is expected as it was only trained with noisy data in continual training and some information learned from the clean data was forgotten. By adding the reconstruction module, our proposed model maintained the performance for clean speech, meanwhile, performed significantly better for noisy speech. With the 4-gram language model applied, our proposed model improved the WERs from 20.9\%, 40.0\% to 7.5\%, 20.2\% for \texttt{test-clean} and \texttt{test-other}, respectively. This suggests that our approach is beneficial for improving the noise robustness of the pre-trained model.

\subsection{Results on Real-world Noisy Speech}

\begin{table}[!tp]
\caption{Experimental results on the real-world noisy speech of the CHiME-4 corpus. We evaluate on the 1-channel track \texttt{dev-real} and \texttt{test-real} subset and calculate the average of the obtained WERs. The first four rows are the results of supervised approaches, and the rest rows are the results obtained using pre-trained self-supervised models that are fine-tuned (FT) with different labelled data.}
\vspace{-.7em}
\resizebox{0.49\textwidth}{!}{\begin{tabular}{@{}cccc@{}}
\toprule
\textbf{ Model}                          &  \textbf{Dev-real} & \textbf{Test-real} & \textbf{Avg.} \\ \midrule
 DNN baseline \cite{vincent2016chime4}   &  11.6 & 23.7 & 17.7  \\
 Du et al. \cite{du2016ustc}             &  4.5  & 9.2  & 6.9  \\
 Menne et al. \cite{menne2016rwth}       &  5.1  & 9.3  & 7.2  \\ 
 Wang et al. \cite{wang2020complex}      &  3.5  & 6.8  & 5.2  \\ \midrule
 Pre-trained (LS960h) FT. W/ 5ch labeled  &  5.7  & 10.7 & 8.2 \\
 Proposed (LS960h) FT. W/ 5ch labeled    &  \textbf{5.0}  & \textbf{9.0}  & \textbf{7.0}  \\
 Proposed (LV60k) FT. W/ 5\%  labeled               &  3.9 & 9.1 & 6.5  \\
 Proposed (LV60k) FT. W/ 16\% labeled               &  3.1 & 7.2 & 5.2  \\
 Proposed (LV60k) FT. W/ 50\% labeled               &  2.8 & 5.8 & 4.3  \\
 Proposed (LV60k) FT. W/ 100\% labeled              &  \textbf{2.7} & \textbf{5.5} & \textbf{4.1} \\
 Pre-trained (LV60k) FT. W/ 100\% labeled      &  2.8 & 5.8 & 4.3 \\ \bottomrule
\end{tabular}}
\label{tbl:chime}

\vspace{-0.2em}
\end{table}

As demonstrated in Table \ref{tbl:chime}, we report evaluation results on the 1-channel real-world noisy speech extracted from the CHiME-4 challenge. 
Two categories of results are displayed, the upper part lists the results of supervised methods. The first three rows are the results reported in the challenge \cite{vincent2016chime4, du2016ustc, menne2016rwth}, and the fourth row displays the best result on the 1-channel track data to the best of our knowledge \cite{wang2020complex}. The lower part of the table lists the results obtained from the pre-trained self-supervised models that are fine-tuned on different percentages of labeled data. 
The major difference between these two categories is that our proposed method does not perform any form of preprocessing, and not all labeled data are required to train the acoustic model. Meanwhile, the supervised approaches conduct speech enhancement to preprocess the noisy inputs before feeding them to acoustic backends. Moreover, to obtain better WERs, techniques like speaker adaptation and ensemble acoustic modeling were employed in supervised studies \cite{menne2016rwth,wang2020complex}.
After applying an LSTM-based language model, we achieved 6.5\% average WER with only 5\% of labeled data, which was 0.4\% better compared with the top-1 result reported in the challenge \cite{du2016ustc}. Furthermore, by using all labeled data for fine-tuning, we reached a 4.1\% WER on average, which was 21.1\% relatively lower than the best WER obtained by the supervised approach \cite{wang2020complex}.
Compared to only utilizing the pre-trained model (no continual training), adding the reconstruction module resulted in around 4.7\% relative WER improvements.
We also show the results using the LibriSpeech (LS960h) based pre-trained model fine-tuned on the five channels of the labeled data (excluding the microphone channel that faces backward). For this relatively small sized pre-trained model, we achieved competitive performance with an average WER of 7.0\%. In addition, adding the reconstruction module to the pre-trained model brought around 14.6\% relative WER improvements on average. The performance boost for the LibriSpeech based model was more obvious, and is likely caused by the difference in the model scale and the size of the training corpus (960 hours versus 60k hours).
This result attests to the potential of our approach, as we are capable of achieving noise robustness with a limited amount of labeled speech and no preprocessing.


\subsection{Ablation Study}
\begin{table}[!tp]
\caption{Ablation study of the proposed architecture (no language model applied). The first two rows investigate the effect of reconstruction module position, and the last row shows the result of replacing the CRN with a BLSTM network.}
\vspace{-.7em}
\resizebox{0.48\textwidth}{!}{\begin{tabular}{@{}ccccc@{}}
\toprule
 \textbf{Model} & \multicolumn{2}{c}{\textbf{Clean}} & \multicolumn{2}{c}{\textbf{Noisy}} \\ \midrule
                &      test-clean & test-other       & test-clean & test-other \\ \midrule
 Proposed       &   10.5          & 19.5             & 14.6       & 29.8     \\ \midrule
 BeforeQuantization       &   10.6          & 19.6             & 14.8       & 30.6   \\
 AfterQuantization        &   10.8          & 20.4             & 14.9       & 30.8   \\
 BLSTM          &   10.2          & 19.5             & 15.6       & 31.3    \\ \bottomrule
\end{tabular}}

\label{tbl:abalation}
\vspace{-0.5em}

\end{table}

To further investigate the effectiveness of our proposed design, we compare variants of the proposed method where the reconstruction module is attached upon different representations.
In addition to the proposed one, we also performed the reconstruction based on the latent representation (after the CNN feature extractor) and the quantized representations (after the quantizer).
Table \ref{tbl:abalation} reports that the proposed method had the overall best performance.
Applying the reconstruction module on the quantized representation (denoted as AfterQuantization) showed the worst WER among the variants, and attaching it before the quantization process (denoted as beforeQuantization) performed slightly better but was not optimal. Under both clean and noisy conditions, the AfterQuantization variant introduced around 1\% WER increment for \texttt{test-other}.
This is likely caused by the difficulties of restoring the clean audio from discretized embeddings.
Additionally, it might be due to the fact that the outputs of the transformer blocks are closer to the final output, and imposing our reconstruction loss upon it has a more direct impact on later acoustic training. 
Furthermore, we have investigated another architecture for the reconstruction module by replacing CRN with a three-layer BLSTM. As shown in the table, under clean conditions, they performed similarly, except that BLSTM was slightly better for \texttt{test-clean}, but under noisy conditions CRN was superior and had a 1.5\% WER reduction for \texttt{test-other}.

\vspace{-0.3em}
\section{Conclusion}
\vspace{-0.1em}

\label{sec:conclusion}
In this paper, we proposed a novel method to improve the noise robustness for self-supervised speech representation learning, which was demonstrated to be effective for robust automatic speech recognition (ASR).
The proposed method combined a reconstruction module with the contrastive learning framework.
Our pre-trained model was continuously trained on noisy/clean speech pairs, then fine-tuned on low-resource labeled data.
Our experiments showed the superiority of the proposed method. The results on both real and synthetic noisy speech showed that adding a reconstruction task during continual training improved the noise robustness of the learned representation.
The limitations of the proposed architecture are that the reconstruction process only imposes implicit constraints on the speech representations, and paired data are required in the continual pre-training stage.
Our future work aims to mitigate this issue by introducing a masker layer for the intermediate representations and incorporating the mixture invariant training \cite{wisdom2020unsupervised} so that no paired data is needed. 
\vfill\pagebreak

\clearpage

\bibliographystyle{IEEEbib}
\bibliography{strings,refs}

\begin{thebibliography}{10}

\bibitem{haeb2019speech}
R.~Haeb-Umbach, S.~Watanabe, T.~Nakatani, M.~Bacchiani, B.~Hoffmeister, M.~L.
  Seltzer, H.~Zen, and M.~Souden,
\newblock ``Speech processing for digital home assistants: Combining signal
  processing with deep-learning techniques,''
\newblock {\em IEEE Signal Processing Magazine}, vol. 36, pp. 111--124, 2019.

\bibitem{wang2018overview}
D.~L. Wang and J.~Chen,
\newblock ``Supervised speech separation based on deep learning: {An}
  overview,''
\newblock {\em IEEE/ACM Transactions on Audio, Speech, and Language
  Processing}, vol. 26, pp. 1702--1726, 2018.

\bibitem{hu2020dccrn}
Y.~Hu, Y.~Liu, S.~Lv, M.~Xing, S.~Zhang, Y.~Fu, J.~Wu, B.~Zhang, and L.~Xie,
\newblock ``{DCCRN}: Deep complex convolution recurrent network for phase-aware
  speech enhancement,''
\newblock in {\em Proceedings of INTERSPEECH}, 2020, pp. 2482--2486.

\bibitem{wang2020complex}
Z.-Q. Wang, P.~Wang, and D.~L. Wang,
\newblock ``Complex spectral mapping for single-and multi-channel speech
  enhancement and robust {ASR},''
\newblock {\em IEEE/ACM transactions on audio, speech, and language
  processing}, vol. 28, pp. 1778--1787, 2020.

\bibitem{meng2017deep}
Z.~Meng, S.~Watanabe, J.~R. Hershey, and H.~Erdogan,
\newblock ``Deep long short-term memory adaptive beamforming networks for
  multichannel robust speech recognition,''
\newblock in {\em Proceedings of ICASSP}, 2017, pp. 271--275.

\bibitem{eskimez2021human}
S.~E. Eskimez, X.~Wang, M.~Tang, H.~Yang, Z.~Zhu, Z.~Chen, H.~Wang, and
  T.~Yoshioka,
\newblock ``Human listening and live captioning: Multi-task training for speech
  enhancement,''
\newblock in {\em Proceedings of INTERSPEECH}, 2021, pp. 2686--2690.

\bibitem{subramanian2019speech}
A.~S. Subramanian, X.~Wang, M.~K. Baskar, S.~Watanabe, T.~Taniguchi, D.~Tran,
  and Y.~Fujita,
\newblock ``Speech enhancement using end-to-end speech recognition
  objectives,''
\newblock in {\em Proceedings of WASPAA}, 2019, pp. 234--238.

\bibitem{watanabe2020ESPupdate}
S.~Watanabe, F.~Boyer, X.~Chang, P.~Guo, T.~Hayashi, Y.~Higuchi, T.~Hori, W-C.
  Huang, H.~Inaguma, N.~Kamo, et~al.,
\newblock ``The 2020 {ESPNet} update: New features, broadened applications,
  performance improvements, and future plans,''
\newblock in {\em Proceedings of DSLW}, 2020, pp. 1778--1787.

\bibitem{wang2021exploring}
X.~Wang, N.~Kanda, Y.~Gaur, Z.~Chen, Z.~Meng, and T.~Yoshioka,
\newblock ``Exploring end-to-end multi-channel {ASR} with bias information for
  meeting transcription,''
\newblock in {\em Workshop of SLT}, 2021, pp. 833--840.

\bibitem{alvarez2019pase}
D.~{\'A}lvarez, S.~Pascual, and A.~Bonafonte,
\newblock ``Problem-agnostic speech embeddings for multi-speaker text-to-speech
  with {SampleRNN},''
\newblock in {\em Proceedings of ISCA Speech Synthesis Workshop}, 2019, pp.
  35--39.

\bibitem{chung2019apc}
Y.-A. Chung, W.-N. Hsu, H.~Tang, and J.~Glass,
\newblock ``An unsupervised autoregressive model for speech representation
  learning,''
\newblock {\em Proceedings of INTERSPEECH}, pp. 146--150, 2019.

\bibitem{liu2020mockingjay}
A.~T. Liu, S.-W. Yang, P.-H. Chi, P.-C. Hsu, and H.-Y. Lee,
\newblock ``Mockingjay: Unsupervised speech representation learning with deep
  bidirectional transformer encoders,''
\newblock in {\em Proceedings of ICASSP}, 2020, pp. 6419--6423.

\bibitem{ling2020bertphone}
S.~Ling, J.~Salazar, Y.~Liu, K.~Kirchhoff, and AWS Amazon,
\newblock ``{BERTphone}: Phonetically-aware encoder representations for
  utterance-level speaker and language recognition,''
\newblock in {\em Proceedings of Odyssey}, 2020, pp. 9--16.

\bibitem{baevski2020wav2vec}
A.~Baevski, H.~Zhou, A.~Mohamed, and M.~Auli,
\newblock ``wav2vec 2.0: A framework for self-supervised learning of speech
  representations,''
\newblock {\em Advances in Neural Information Processing Systems}, vol. 33,
  2020.

\bibitem{oord2018cpc}
A.~V. Oord, Y.~Li, and O.~Vinyals,
\newblock ``Representation learning with contrastive predictive coding,''
\newblock {\em arXiv:1807.03748}, 2018.

\bibitem{baevski2019vq}
A.~Baevski, S.~Schneider, and M.~Auli,
\newblock ``vq-wav2vec: Self-supervised learning of discrete speech
  representations,''
\newblock {\em arXiv:1910.05453}, 2019.

\bibitem{kawakami2020learning}
K.~Kawakami, L.~Wang, C.~Dyer, P.~Blunsom, and A.~Oord,
\newblock ``Learning robust and multilingual speech representations,''
\newblock in {\em Proceedings of EMNLP}, 2020, pp. 1182--1192.

\bibitem{wang2021unispeech}
C.~Wang, Y.~Wu, Y.~Qian, K.~Kumatani, S.~Liu, F.~Wei, M.~Zeng, and X.~Huang,
\newblock ``Unispeech: Unified speech representation learning with labeled and
  unlabeled data,''
\newblock in {\em Proceedings of ICML}, 2021, pp. 10937--10947.

\bibitem{ravanelli2020paseplus}
M.~Ravanelli, J.~Zhong, S.~Pascual, P.~Swietojanski, J.~Monteiro, J.~Trmal, and
  Y.~Bengio,
\newblock ``Multi-task self-supervised learning for robust speech
  recognition,''
\newblock in {\em Proceedings of ICASSP}, 2020, pp. 6989--6993.

\bibitem{sadhu2021wav2vecc}
S.~Sadhu, D.~He, C.-W. Huang, S.~H. Mallidi, M.~Wu, A.~Rastrow, A.~Stolcke,
  J.~Droppo, and R.~Maas,
\newblock ``{Wav2vec-C}: A self-supervised model for speech representation
  learning,''
\newblock in {\em Proceedings of INTERSPEECH}, 2021, pp. 711--715.

\bibitem{tan2018complex}
K.~Tan and D.~L. Wang,
\newblock ``A convolutional recurrent neural network for real-time speech
  enhancement,''
\newblock in {\em Proceedings of INTERSPEECH}, 2018, pp. 3229--3233.

\bibitem{panayotov2015librispeech}
V.~Panayotov, G.~Chen, D.~Povey, and S.~Hudanpur,
\newblock ``Librispeech: an {ASR} corpus based on public domain audio books,''
\newblock in {\em Proceedings of ICASSP}, 2015, pp. 5206--5210.

\bibitem{snyder2015musan}
D.~Snyder, G.~Chen, and D.~Povey,
\newblock ``{MUSAN}: A music, speech, and noise corpus,''
\newblock {\em arXiv:1510.08484}, 2015.

\bibitem{vincent2016chime4}
E.~Vincent, S.~Watanabe, J.~Barker, and R.~Marxer,
\newblock ``The 4th {CHiME} speech separation and recognition challenge,''
\newblock {\em URL: http://spandh. dcs. shef. ac. uk/chime\_challenge/}, 2016.

\bibitem{paul1992wsj}
D.~B. Paul and J.~M. Baker,
\newblock ``The design for the wall street journal-based {CSR} corpus,''
\newblock in {\em Proceedings of a Workshop on Speech and Natural Language},
  1992.

\bibitem{kahn2020librilight}
J.~Kahn, M.~Rivi{\`e}re, W.~Zheng, E.~Kharitonov, Q.~Xu, P.-E. Mazar{\'e},
  J.~Karadayi, V.~Liptchinsky, R.~Collobert, C.~Fuegen, et~al.,
\newblock ``Libri-light: A benchmark for {ASR} with limited or no
  supervision,''
\newblock in {\em Proceedings of ICASSP}, 2020, pp. 7669--7673.

\bibitem{heafield2011kenlm}
K.~Heafield,
\newblock ``{KenLM}: Faster and smaller language model queries,''
\newblock in {\em Proceedings of the statistical machine translation}, 2011,
  pp. 187--197.

\bibitem{wang2019espresso}
Y.~Wang, T.~Chen, H.~Xu, S.~Ding, H.~Lv, Y.~Shao, N.~Peng, L.~Xie, S.~Watanabe,
  and S.~Khudanpur,
\newblock ``Espresso: A fast end-to-end neural speech recognition toolkit,''
\newblock in {\em Proceedings of ASRU}, 2019, pp. 136--143.

\bibitem{du2016ustc}
J.~Du, Y.-H. Tu, L.~Sun, F.~Ma, H.-K. Wang, J.~Pan, C.~Liu, J.-D. Chen, and
  C.-H. Lee,
\newblock ``The {USTC-iFlytek} system for {CHiME}-4 challenge,''
\newblock {\em Proc. CHiME}, vol. 4, pp. 36--38, 2016.

\bibitem{menne2016rwth}
T.~Menne, J.~Heymann, A.~Alexandridis, K.~Irie, A.~Zeyer, M.~Kitza, P.~Golik,
  I.~Kulikov, L.~Durde, R.~Schl{\"u}ter, H.~Ney, R.~Haeb-Umbach, and
  A.~Mouchtaris,
\newblock ``The rwth/upb/forth system combination for the 4th chime challenge
  evaluation,''
\newblock {\em Computer Speech \& Language}, 2016.

\bibitem{wisdom2020unsupervised}
S.~Wisdom, E.~Tzinis, H.~Erdogan, R.~J. Weiss, K.~Wilson, and J.~R. Hershey,
\newblock ``Unsupervised sound separation using mixture invariant training,''
\newblock in {\em Proceedings of NeurIPS}, 2020.

\end{thebibliography}

\end{document}